\documentclass[12pt]{article}
\usepackage{graphicx,psfrag}
\newcommand{\preprint}[1]{\begin{flushright}#1\end{flushright}}

\newcommand{\bea}{\begin{eqnarray}}
\newcommand{\eea}{\end{eqnarray}}
\newcommand{\simgt}{\hbox{ \raise3pt\hbox to 0pt{$>$}\raise-3pt\hbox{$\sim$} }}
\newcommand{\simlt}{\hbox{ \raise3pt\hbox to 0pt{$<$}\raise-3pt\hbox{$\sim$} }}

\def\to{\rightarrow}

\begin{document}
\preprint{TU-636\\October 2001}
\vspace*{3cm}
\begin{center}
  {\large
Off-shell Suppression of Renormalons in 
Non-relativistic QCD Boundstates}
  \\[15mm]
  {\large
    Y.~Kiyo and Y. Sumino
    }
  \\[10mm]
  {\it
    Department of Physics, Tohoku University\\
    Sendai, 980-8578 Japan
    }
\end{center}
\vspace{25mm}
\begin{abstract}
We study an effect 
of the off-shellness of the quark and antiquark
inside a heavy quarkonium system on IR renormalons contained
in the perturbative computations of the quarkonium energy levels.
We demonstrate that, when the off-shellness 
$p_Q^2 -m_Q^2 \sim \alpha_S^2 m_Q^2$
is larger than $m_Q \Lambda_{\rm QCD}$, 
renormalons in the energy levels as calculated in perturbative QCD are
suppressed by a factor
$\Lambda_{\rm QCD}/(\alpha_S^2 m_Q)$
as compared to those in 
$2 m_{\rm pole} + V_{\rm QCD}(r)$.
In this case the residual
${\cal O}(\Lambda_{\rm QCD}^4)$ renormalon has the same dimension as
that of the leading gluon-condensate contribution.
\end{abstract}

\newpage

Recently, there has been renewed interest in 
infrared (IR) renormalons\footnote{ 
See \cite{Beneke99} for a review of renormalons.
}
in the context of perturbative
computations of the physical quantities of the heavy quarkonium states,
within the framework of effective theories
such as Non-Relativistic QCD (NRQCD) \cite{NRQFT}
and potential-NRQCD (pNRQCD) \cite{pNRQCD}.
This is because, proper identification of renormalons and
realization of their cancellations in these perturbative computations 
have led to significant improvements in the determinations of the heavy 
quark masses $m_t$, $m_b$ and $m_c$ 
\cite{TopWG,Pineda,KiyoSumino,Hoang1,BraSuminoVai1,BraSuminoVai2}, 
as well as in the perturbative computations of the heavy quarkonium
spectra \cite{BraSuminoVai1,BraSuminoVai2} and the static QCD potential
\cite{Sumino01,ReckSumino}.
We have learned that through a deeper understanding of renormalons 
in these systems we may achieve
more accurate perturbative predictions, and consequently 
we may gain more accurate
physical picture of the heavy quarkonium states.
So far, our understanding of IR renormalons in the perturbative
computations has been as follows. 
(See also the discussion at the end of the paper.)
For simplicity we consider the case where the  
quark $Q$ and antiquark $\bar{Q}$ have equal masses in this paper.
When the total energy of a static $Q\bar{Q}$ pair
$E_{\rm tot}(r) \equiv 2m_{\rm pole}+V_{\rm QCD}(r)$
is expressed in terms of a short-distance mass of $Q$($\bar{Q}$),
where $m_{\rm pole}$ is the pole mass of $Q$($\bar{Q}$) and
$V_{\rm QCD}(r)$ is the static QCD potential,
the leading renormalon of order $\Lambda_{\rm QCD}$ 
contained in $2 m_{\rm pole}$ \cite{BenekeBraun94,BSUV}
(the self-energies of $Q$ and $\bar{Q}$)
is cancelled \cite{HSSW,Beneke98}
against that contained in $V_{\rm QCD}(r)$ \cite{AgLig}
(the potential energy between $Q$ and $\bar{Q}$).
Remaining sub-leading renormalons in $E_{\rm tot}(r)$ are of order
$\Lambda_{\rm QCD}\times(r\Lambda_{\rm QCD})^2$ \cite{AgLig,BraPineSotVai},
where $r$ is the size of the boundstate.
Hence, they are smaller than the leading renormalon contributions
for a heavy quarkonium system, for which 
$r \Lambda_{\rm QCD} \ll 1$.
As a result, the perturbative expansion of the total energy
$E_{\rm tot}$ becomes much more convergent when the expansion is
expressed in terms of a short-distance mass than when
expressed in terms of the pole mass.

In perturbative QCD calculations, 
the order $\Lambda_{\rm QCD}$ leading renormalons
may be identified as follows
\cite{Beneke98}.
The pole mass and the
QCD potential constitute the total energy of a $Q\bar{Q}$ system in which
$Q$ and $\bar{Q}$ are on-shell.
Dominant contributions from IR regions to these quantities
can be written (independently of the quark short-distance mass) as
\begin{eqnarray}
&&
2m_{\rm pole}\sim
-V_{\rm QCD}
\sim 
\int \frac{ d^3 \vec{k}
         }{ \left(2\pi\right)^3 } 
\frac{ \alpha_S(|\vec{k}|)
    }{ |\vec{k}|^2} .
\end{eqnarray}
The integral is dominated by contributions from
$|\vec{k}|\sim \Lambda_{\rm QCD}$
where $\alpha_S(|\vec{k}|) $ becomes large.

The above argument, however, needs to be modified in the case where $Q$ and 
$\bar{Q}$ are persistently off-shell.
In a realistic quarkonium system, $Q$ and $\bar{Q}$ are off-shell
due to the binding energy.
The off-shellness in a sufficiently heavy quarkonium system
is given by 
$p_Q^2 -m^2 \sim {\cal O}(\alpha_{s}^2 m^2)$.
Throughout this paper, $m$ represents the short-distance mass of
$Q$ and $\bar{Q}$, which
differs from the pole mass (formally) by an amount of
order $\alpha_{s}^2 m$ or smaller within perturbative QCD,
e.g.\ the Potential-Subtracted mass
with an appropriate choice of the subtraction scale \cite{Beneke98}.

Let us consider the time evolution of a quark-antiquark pair bound in a heavy
quarkonium state.
The potential $V_{\rm QCD}(r)$
generated by a static color source contains a renormalon of
${\cal O}(\Lambda_{\rm QCD})$.
We may infer that this feature is related to the fact that 
vacuum polarization by gluons is induced around the bare
color source,
and that an energy of ${\cal O}(\Lambda_{\rm QCD})$ is accumulated
due to the antiscreening effects.
When the quark  (color source) is static, 
the antiscreening would be completed in a time scale
$\Delta t\sim \Lambda_{\rm QCD}^{-1}$
needed for the gluon clouds to surround the quark.
This picture is consistent with the observation \cite{BSUV,Beneke98}
that the leading renormalons
contained in the pole mass and the static QCD potential,
which are associated with on-shell quarks, stem from contributions of
instantaneous gluons (in the quark rest frame, and independently of gauge
choice).
Since the on-shell quark is static (stable for infinite time),
we may regard the time scale 
$\Delta t\sim \Lambda_{\rm QCD}^{-1}$
as instantaneous comparatively.
On the other hand, in a realistic heavy quarkonium state, 
$Q$ and $\bar{Q}$ are off-shell.
This means that $Q$ and $\bar{Q}$ are not static color sources but 
they keep rescattering with each other at a time interval
$\Delta t\sim 1/(\alpha_S^2 m)$.
Therefore, if the off-shellness is large
$1/\alpha_S^2 m \ll 1/\Lambda_{\rm QCD}$,
$Q$ or $\bar{Q}$ at rest (in one frame) 
would be scattered before the antiscreening is completed,
and the energy accumulated around it would be smaller than
${\cal O}(\Lambda_{\rm QCD})$.

\begin{figure}
\begin{center}
\psfrag{kx}{$~~~p_Q$}  
\psfrag{kxx}{$\!\!\!\!\! p_Q \! +k$}
\psfrag{kxxx}{$k$}
\psfrag{pixx}{$\Pi(k^2)$}
\psfrag{Qt}{$Q$}
\psfrag{Qtbar}{$\bar{Q}$}
\includegraphics[width=6cm]{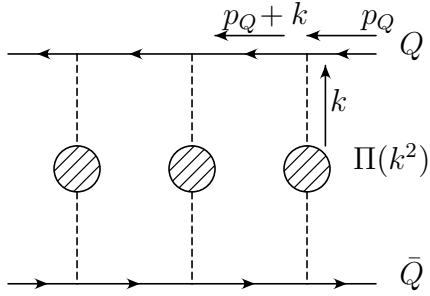}
\end{center}
\vspace*{-.5cm}
\caption{\footnotesize
A diagram showing typical propagation of 
$Q$ and $\bar{Q}$ in a heavy quarkonium state.}
\label{propagation}
\end{figure}

Fig.~\ref{propagation} shows typical propagation of $Q$ and $\bar{Q}$ in a 
heavy quarkonium state.
In the leading kinematical configuration contributing to the formation of the
non-relativistic boundstate, gluons exchanged
between $Q$ and $\bar{Q}$ have momenta
$ k^0\sim {\cal O}(\alpha_S^2 m),~ |\vec{k}|\sim {\cal
O}(\alpha_S m)$.
In an approximation valid for this configuration,
we may replace the gluon vacuum polarization as
$\Pi(k^2) \simeq \Pi(-|\vec{k}|^2) $.
Then we may easily integrate over the time-component of the 
loop momenta, and the integral is dominated by contributions
from kinematical regions where $Q$ and $\bar{Q}$ are nearly on-shell.
On the other hand, it 
is not obvious whether the same approximation is valid also 
in estimating IR renormalons,
since a different kinematical region, $-k^2 \sim \Lambda_{\rm QCD}^2$,
contributes to them.
In fact, two scales $k^0\sim\Lambda_{\rm QCD}$ and 
$p_Q^0-m-{\vec{p}_Q^{\,2}}/{2m}\sim\alpha_S^2m$
compete in the quark propagator
when $p_Q^2-m^2\sim \alpha_S^2 m^2$ and
$k^0 ,|\vec{k}| \sim \Lambda_{\rm QCD}$,
since $(p_Q+k)^2-m^2\simeq 2m
\left[p_Q^0+k^0-m-{(\vec{p}_Q+\vec{k})^2}/{2m}\right]$.
Furthermore, when
$p_Q^2-m^2\sim \alpha_S^2 m^2$ and $(p_Q+k)^2 - m^2 \approx 0$,
it follows that $k^0\sim \alpha_S^2 m$,
$|\vec{k}|\sim \alpha_S m$, i.e.\ $k^2 \sim \alpha_S^2 m^2$,
hence in this (nearly on-shell) configuration $\Pi(k^2)$ is small and 
does not contribute to renormalons.
In particular, in estimating renormalon contributions by power counting  
in the loop integral of the diagram of Fig.~\ref{propagation},
we should count 
\begin{eqnarray}
&&
\int\frac{dk^0}{2\pi} \, \frac{m}{(p_Q+k)^2-m^2}
\sim
\frac{\Lambda_{\rm QCD}}{\alpha_S^2 m} ,
\label{powercount}
\end{eqnarray}
if $\alpha_S^2 m \gg \Lambda_{\rm QCD}$.
This order estimate is suppressed as compared to the estimate
in the on-shell approximation of ${\cal O}(1)$.
According to the power counting (\ref{powercount}),
the leading renormalon contained in the potential energy between
$Q$ and $\bar{Q}$ is estimated to be of order
$ \Lambda_{\rm QCD} \times (\Lambda_{\rm QCD}/(\alpha_S^2 m))$
instead of order $\Lambda_{\rm QCD}$.
Similarly, the leading renormalon contained in the self-energy of
$Q$ or $\bar{Q}$ is of order
$\Lambda_{\rm QCD} \times (\Lambda_{\rm QCD}/(\alpha_S^2 m))$
and is suppressed 
in comparison to the renormalon of order $\Lambda_{\rm QCD}$
in the pole mass, if we take into account the off-shellness.
We will show the power countings more explicitly in our discussion below.

The above argument suggests that, in order to estimate renormalon 
contributions to the heavy quarkonium state accurately in the case
$\alpha_S^2 m \gg \Lambda_{\rm QCD}$, we should make an approximation 
valid both in the leading kinematical configuration 
of the boundstate and in the
renormalon configuration.
Estimates of renormalons within an effective theory
may in general
differ from the renormalons of perturbative QCD, if the effective
theory fails to
incorporate the off-shellness appropriately.
Therefore, in this paper, we start from the Bethe-Salpeter (BS) formalism
(see e.g.\ \cite{BodwinYennie}), where one may incorporate (in
principle) all the contributions of perturbative QCD,
in estimating renormalons in the energy levels of boundstate.
According to the BS formalism, the potential energy 
$E_{\rm pot}$ 
and the self-energy contributions $E_{\rm SE}$ in the
total energy of the boundstate may be expressed,
in the rest frame of the boundstate, as
\begin{eqnarray}
&&
E_{\rm pot}
=
\frac{i}{2M} \,
\left(
\overline{\chi} \cdot K \cdot \chi 
\right)
~,
\\
&&
E_{\rm SE}
=
M -
\frac{i}{2M} \,
\left(
\overline{\chi} \cdot  \left[
(S_{F,Q})^{-1}(S_{F,\bar{Q}})^{-1} \right]
\cdot \chi 
\right)
~.
\end{eqnarray}
Here, $\chi,\overline{\chi}$ and $K$ denote the BS wave functions
\cite{LuMacTak}
and the BS-kernel, respectively; 
$M$ is the boundstate mass (total energy);
$S_F$ represents the full propagator of $Q$ or $\bar{Q}$.
We set the momentum of the center of gravity as
$(M,\vec{0})$.
The dot ($\cdot$) represents contraction of spinor indices
and an integral over the relative momentum 
between $Q$ and $\bar{Q}$.
Diagrammatically,
$E_{\rm pot}$ ($E_{\rm SE}$) represents the contributions from the diagrams
where the $Q$ and $\bar{Q}$ lines
are connected (disconnected), and
$E_{\rm pot}+E_{\rm SE}=M$.

As an example, we estimate renormalons contained in the potential energy
$E_{\rm pot}$ of the $1S$ state.
Following the standard procedure, we replace the BS-kernel $K$ by the 
Coulomb propagator (in Coulomb gauge) in the
large-$\beta_0$ approximation \cite{BenekeBraun95}:
\bea
K \to K_{\beta_0}(k) =
-i \, \frac{ C_F \, 4 \pi\alpha_S(\mu)}{|\vec{k}|^2} \,
( \gamma^0 \otimes \gamma^0 ) \times
\sum_{n=0}^\infty \left( \frac{\beta_0 \alpha_S(\mu)}{4\pi} \right)^n
\left\{ \log \Biggl( \frac{\tilde{\mu}^2}{-k^2} \Biggr) \right\}^n .
\eea
Here, $\alpha_S(\mu)$ is the coupling constant in the
$\overline{\rm MS}$ scheme, and 
$\mu$ is the renormalization scale.
$\beta_0=11-\frac{2}{3}n_l$ denotes the one-loop coefficient of the
beta function with $n_l$ flavors;
$\tilde{\mu}=e^{5/6} \mu$; $C_F=4/3$ is the color factor.
Similarly, it would be desirable to replace the BS wave functions
$\chi(p),\overline{\chi}(p)$ by those in the large-$\beta_0$ approximations.
Since, however, they cannot be obtained simply, in our calculation
we substitute the leading-order wave functions in $1/c$ expansion:
\bea
&&
\chi(p) \to
\chi^{(0)}_{1S} (p)=
\left( \frac{i}{\frac{M}{2}+p^0-m-\frac{\vec{p}^{\,2}}{2m}+i\epsilon}
      +\frac{i}{\frac{M}{2}-p^0-m-\frac{\vec{p}^{\,2}}{2m}+i\epsilon}
\right) 
\sqrt{2M} \, \phi_{C,1S}(\vec{p}) \times \lambda  ,
\nonumber \\
&&
\overline{\chi}(p) \to
\overline{\chi}^{(0)}_{1S} (p)=
\left( \frac{i}{\frac{M}{2}+p^0-m-\frac{\vec{p}^{\,2}}{2m}+i\epsilon}
      +\frac{i}{\frac{M}{2}-p^0-m-\frac{\vec{p}^{\,2}}{2m}+i\epsilon}
\right) 
\sqrt{2M} \, \phi_{C,1S}^*(\vec{p}) \times \lambda^\dagger .
\label{chi0}
\nonumber \\
\eea
Here, $ \phi_{C,1S}(\vec{p})$ and $M$ denote the
$1S$ Coulomb wave function and its energy level, respectively:
\begin{eqnarray}
&&
\phi_{C,1S}(\vec{p})
=
\frac{ \sqrt{2\pi} \left(C_F \alpha_S(\mu) m \right)^{ \frac{5}{2}}
    }
{ \left[ \vec{p}^{\,\, 2} + \left( C_F \alpha_S(\mu) m/2\right)^2 \right]^2}
~,
~~~~~~~~~~
M = 2 \, m - \frac{\Bigl( C_F \alpha_S(\mu) \Bigr)^2}{4} \, m .
\end{eqnarray}
$\lambda$ is a spinor matrix representing the
spin of the boundstate,
\begin{eqnarray}
&&
\lambda_{J=1}
=
\frac{1+\gamma^0}{2}\, 
\frac{\varepsilon ~\hspace{-9pt}\slash}{\sqrt{2}}\,
\frac{1-\gamma^0}{2} ,
~~~~~~~
\lambda_{J=0}
=
\frac{1+\gamma^0}{2} \,
\frac{\gamma_5}{\sqrt{2}}\,
\frac{1-\gamma^0}{2} ,
\end{eqnarray}
with polarization vector $\varepsilon$ for the vector $1S$ state.
Power counting shows that, as far as the dependence on the
wave functions $\chi(p)$,$\bar{\chi}(p)$ is concerned,
the order estimates of renormalons are determined by the
analyticity of $\chi(p)$,$\bar{\chi}(p)$ and by
the support in 
momentum-space of the part corresponding to $\phi_{C,1S}(\vec{p})$.
Therefore, we conjecture that our approximation 
for the wave functions (\ref{chi0}) 
does not alter order estimates of renormalons.

In these approximations, the potential energy is given by
\bea
E_{{\rm pot},\beta_0}^{(1S)} = 
\frac{i}{2M} \,
\int \frac{d^4p}{(2\pi)^4}
\int \frac{d^4k}{(2\pi)^4}
\, \, \, 
\overline{\chi}^{(0)}_{1S}(p) 
\, \, K_{\beta_0}(k) \, \, \chi^{(0)}_{1S}(p+k) .
\label{potene}
\eea
Its Borel transform reads
\bea
&&
\widetilde{E}_{{\rm pot},\beta_0}^{(1S)} [u] = 
- \, \frac{\Bigl( C_F \alpha_S(\mu) \Bigr)^2 m}{2} \times
\left( \frac{\tilde{\mu}}{C_F \alpha_S(\mu) m} \right)^{2u} 
{I}(u,\Delta) ~,
\label{Borelpot}
\\
&&
{I}(u,\Delta) \equiv
- \, 2^{8+2u} \,i \, \pi^2
\int \! \frac{d^3\vec{p}}{(2\pi)^3}
\int \! \frac{d^4k}{(2\pi)^4}
\, \, \, 
\frac{1}{
\Bigl( \vec{p}^{\,\,2}+1 \Bigr)^2 \,
\Bigl( | \vec{p} + \vec{k} |^2 +1 \Bigr)^2 \,
\Bigl| \vec{k} \Bigr|^2 \,
\Bigl[ - (k^0 \Delta )^2 + |\vec{k}|^2 - i \epsilon \Bigr]^u
}
\nonumber \\
&&
~~~~~~~ ~~~~~~~~
\times 
\left(
\frac{1}{k^0+1+{\vec{p}^{\,\,2}}/{2}+{|\vec{p}+\vec{k}|^2}/{2}-i\epsilon}
+
\frac{1}{-k^0+1+{\vec{p}^{\,\,2}}/{2}+{|\vec{p}+\vec{k}|^2}/{2}-i\epsilon}
\right) .
~~~
\nonumber\\
\label{I}
\eea
Note that in the integral of ${I}(u,\Delta)$,
we have rescaled the 3-momenta $\vec{p},\vec{k}$
and the energy $k^0$ by the Bohr scale
$p_B=C_F \alpha_S(\mu) m/2$ and the Coulomb binding energy
$E_B=(C_F \alpha_S(\mu))^2 m/4$, respectively 
($p^0$ integration has already been performed).
We have introduced a dimensionless parameter 
$\Delta\equiv E_B/p_B=C_F\alpha_S/2$, which characterizes the
off-shellness of $Q$ and $\bar{Q}$ in the boundstate.
The Borel parameter $u$ is defined with respect to 
$\beta_0 \alpha_S(\mu)/(4\pi)$,
so that the potential energy is given by
\begin{eqnarray}
&&
E_{{\rm pot},\beta_0}^{(1S)}
=
\sum_{n=0}^{\infty}
\left(\frac{\beta_0 \alpha_S}{4\pi}\right)^n
\left[ \frac{\partial^n}{\partial u^n} 
\widetilde{E}_{{\rm pot},\beta_0}^{(1S)} [u] \right]_{u=0} .
\label{PTexpansion}
\end{eqnarray}

If we take the limit $\Delta \to 0$ in ${I}(u,\Delta)$ [Eq.~(\ref{I})]
{\it before} we perform the integration, it can be evaluated easily as
\bea
&&
J(u) \equiv
- \, 2^{8+2u} \, i \, \pi^2
\int \! \frac{d^3\vec{p}}{(2\pi)^3}
\int \! \frac{d^4k}{(2\pi)^4}
\, \, \, 
\frac{1}{
\Bigl( \vec{p}^{\,\,2}+1 \Bigr)^2 \,
\Bigl( | \vec{p} + \vec{k} |^2 +1 \Bigr)^2 \,
\Bigl| \vec{k} \Bigr|^{2+2u} 
}
\nonumber \\
&&
~~~~~~~ ~~~~
\times 
\left(
\frac{1}{k^0+1+{\vec{p}^{\,\,2}}/{2}+{|\vec{p}+\vec{k}|^2}/{2}-i\epsilon}
+
\frac{1}{-k^0+1+{\vec{p}^{\,\,2}}/{2}+{|\vec{p}+\vec{k}|^2}/{2}-i\epsilon}
\right)
~~~
\\
&& ~~~~
=
\frac{1+2u}{\cos (\pi u)} .
\eea
If we replace ${I}(u,\Delta)$ by $J(u)$ in Eq.~(\ref{Borelpot}),
it reduces to the Borel transform of the potential energy calculated
from the static QCD potential in similar approximations.\footnote{
This can be obtained most easily by taking the expectation value
of the Borel transform of the QCD potential in the large-$\beta_0$
approximation \cite{AgLig} with respect to the $1S$ Coulomb wave
function (in position-space).
}
In this case, IR renormalon poles are located at
$u=\frac{1}{2}, \, \frac{3}{2}, \, \frac{5}{2}, \dots$
on the positive real axis.

Using Feynman-parameter integrals, we have reduced ${I}(u,\Delta)$
to a one-parameter integral form including hypergeometric 
functions:
\begin{eqnarray}
&& 
{I}(u,\Delta)
= 
\frac{4 \,
         \Gamma(\frac{5}{2}+u)
         \Gamma(2+2u)
         \Gamma(1-u)
    }{\pi^{3/2}\, \Delta \,
         \Gamma(u)
         \Gamma(3+u)} 
~\int_0^1 d\xi
~f(\xi , u)
{\textstyle
_{~2}F_{1} \! \left(\frac{5}{2}+u,1-u;3+u;1-\frac{\xi}{\Delta^2}\right)
},
\nonumber \\
\label{oneparamint}
&& \\
&&
f(\xi,u)
=
      \frac{1-\xi}{\sqrt{\xi}} ~
{\textstyle
       {_2 F_1} \! \left( -\frac{1}{2}-u,
                       \frac{1}{2};
                        \frac{3}{2};
                        \left(\frac{1-\xi}{1+\xi}\right)^2
                  \right)
}
      -\frac{ 2(1-\xi)}{1+\xi} ~
{\textstyle
      {_2 F_1} \! \left(   -u,
                        \frac{1}{2};
                        \frac{3}{2};
                        \left(\frac{1-\xi}{1+\xi}\right)^2
                  \right) .
\label{fxiu}
}
\end{eqnarray} 
As can be verified easily, the integral in Eq.~(\ref{oneparamint})
is regular at ${\rm Re}\,u > 0$ and exhibits a simple pole as $u \to 0$.
Combining with the prefactor, 
we see that $\widetilde{E}_{{\rm pot},\beta_0}^{(1S)} [u]$
has simple poles only at the positive integers 
$u=N$ $(N=1,2,3,\cdots)$ in the region ${\rm Re}\,u \geq 0$.
The residue $R_N$ of the pole at $u=N$ is a $(2N-1)$-th-degree polynomial of 
$\Delta^{-1}$ and includes only odd powers of $\Delta^{-1}$.
First few residues $R_N$ are listed in Table~\ref{RN}.

\begin{table}
\begin{center}
\begin{tabular}{c|c} \hline
   $N$  & 
$  R_{N}\times \Bigl[ - \, \frac{( C_F \alpha_S )^2 m}{2} \Bigr]^{-1}$ 
\\ \hline \hline
 $1$ &  
$\rule[-3mm]{0mm}{7mm}
x^2 \times 
{\frac{-5}{2\,\pi \,\Delta }}$ \\ \hline
 $2$ &  
$\rule[-3mm]{0mm}{7mm}
 x^4 \times \left(
{   \frac{21}{8\,\pi \,{{\Delta }^3}}} 
 + {\frac{91}{24\,\pi \,\Delta }  }
             \right)$ \\ \hline
 $3$ &  
$\rule[-3mm]{0mm}{7mm}
 x^6 \times \left(
{-{\frac{1287}{256\,\pi \,{{\Delta }^5}}}} 
 -{\frac{429}{128\,\pi \,{{\Delta }^3}}} 
 -{\frac{1467}{256\,\pi \,\Delta } }
             \right)$ \\ \hline
 $4$ &  
$\rule[-3mm]{0mm}{7mm}
 x^8 \times \left(
{\frac{12155}{1024\,\pi \,{{\Delta }^7}}} + {\frac{6149}{1024\,\pi \,{{\Delta }^5}}} + 
  {\frac{26169}{5120\,\pi \,{{\Delta }^3}}} + {\frac{8195}{1024\,\pi \,\Delta }}
             \right)$ \\ \hline
\end{tabular} 
\caption{ \label{RN} \footnotesize
The residues of the poles at $u=N$ in 
$\widetilde{E}_{{\rm pot},\beta_0}^{(1S)} [u]$,
normalized by 
$-\frac{( C_F \alpha_S )^2 m}{2}$;
$x \equiv \tilde{\mu}/(C_F \alpha_S(\mu) m)$.}
\end{center}
\end{table}

The asymptotic behavior of the perturbative expansion of
${E}_{{\rm pot},\beta_0}^{(1S)}$ at large orders
is controlled by the poles at $u=1,2,3,\cdots$.
The leading contribution comes from the pole at $u=1$ and
is given by
\begin{eqnarray}
&&
{E}_{{\rm pot},\beta_0}^{(1S)}
\sim 
\sum_{n=0} 
\left(\frac{\beta_0 \alpha_S}{4\pi}\right)^n 
\left[\frac{\partial^n}{\partial u^n} 
        \left(\frac{R_1}{u-1}
        \right)
\right]_{u=0}
=
- R_1 \times 
\sum_{n=0}^{\infty} \left(\frac{\beta_0\alpha_S}{4\pi}\right)^n n! 
~.
\end{eqnarray}
We may estimate the uncertainty originating from the
leading renormalon pole at $u=1$ as\footnote{
Given the asymptotic series 
\vspace{-2mm}
$\displaystyle 
A=\sum_{n=0}^{\infty} n! \, a^n 
$
with a small parameter $a>0$,
we consider the finite sum 
up to $n=n_\star$,
$\displaystyle 
A_{opt}=\sum_{n=0}^{n_\star} n! \, a^n
$,
to be an approximate value of $A$,
where $n_\star\approx 1/a$ is the order $n$ at which
the term of the asymptotic series becomes smallest.
Since the size of the term barely changes in the range
$n \in ( n_\star - \sqrt{n_\star}, n_\star + \sqrt{n_\star} )$,
we may consider that the estimate by the finite sum has a
truncation error of order
$
\delta A_{opt}
\sim \sqrt{n_\star}\times n_\star ! \, a^{n_\star} 
\sim {n_\star} \, e^{-{n_\star}}
$.
(See e.g.\ \cite{Sumino00}.)
}
\begin{eqnarray}
&&
\delta {E}_{{\rm pot},\beta_0}^{(1S)}
\sim 
\Lambda_{\rm QCD} 
\times 
\left( \frac{ \Lambda_{\rm QCD}
           }{\alpha_S^2 m}
\right) .
\end{eqnarray}
Similarly, the uncertainty generated by the renormalon pole at
$u=N$ may be estimated as
\begin{eqnarray}
&&
\delta {E}_{{\rm pot},\beta_0}^{(1S)}
\sim 
\Lambda_{\rm QCD} \times 
\left(\frac{ \Lambda_{\rm QCD}
          }{ \alpha_S^2 m}
\right)^{2N-1} \times
\left(   f_0
        +f_1 \, \alpha_S^2
        +f_2 \, \alpha_S^4
        +\cdots
        +f_{N-1} \, \alpha_S^{2N-2}
\right) .
\label{subleadingren}
\end{eqnarray}
The polynomial of $\alpha_S(\mu)$ follows from
the form of the residue $R_N$:
$f_{l}$ denotes the coefficient of $\alpha_S^{2l}$
in $(\alpha_S^{2N-2}\, R_{N})$.
We may rearrange the above uncertainties for all $N$ and rewrite them as
\bea
\delta {E}_{{\rm pot},\beta_0}^{(1S)}
\sim 
\Lambda_{\rm QCD} \times
\left(\frac{ \Lambda_{\rm QCD}
          }{ \alpha_S m}
\right)^{2a} \times
\left(\frac{ \Lambda_{\rm QCD}
          }{ \alpha_S^2 m}
\right)^{2b+1} ,
\label{suppressed}
\eea
where $a,b \geq 0$. 
Correspondence to Eq.~(\ref{subleadingren}) can be made via $N=a+b+1$.  
Since
in the on-shell approximation, the pole at $u=a + \frac{1}{2}$ in $J(u)$
induces an uncertainty of order 
$\Lambda_{\rm QCD} \times ({\Lambda_{\rm QCD}}/{(\alpha_S m}))^{2a}$,
we see that each 
renormalon uncertainty is suppressed by factors
$({\Lambda_{\rm QCD}}/{(\alpha_S^2 m}))^{2b+1}$ due to
off-shell effects. 

Let us compare the above result with the order estimates of renormalons
by power counting in the case
$\alpha_S^2 m \gg \Lambda_{\rm QCD}$.
We apply the following counting rules to Eq.~(\ref{potene}):
\bea
k^0, |\vec{k}| \sim \Lambda_{\rm QCD},~~~
p^0 \sim \alpha_S^2 m, ~~~
|\vec{p}| \sim \alpha_S m .
\eea
Also, we regard the kernel as 
$K_{\beta_0}(k) \sim \alpha_S(\sqrt{-k^2})/|\vec{k}|^2$
and count (only) this coupling as 
$ \alpha_S(\sqrt{-k^2}) \sim {\cal O}(1)$.
[For example, we count the integral measures as
$d^4p \sim \alpha_S^5 m^4$, $d^4k \sim \Lambda_{\rm QCD}^4$,
the Coulomb wave function as
$\phi_{C,1S}(\vec{p}) \sim (\alpha_S m )^{-3/2} $, etc.]
We expand the propagators and $\phi_{C,1S}(\vec{p}+\vec{k})$
in terms of $k^0$ and $\vec{k}$ in the integrand.
Due to the symmetries under 
$k^0 \leftrightarrow -k^0$ and $\vec{k} \leftrightarrow -\vec{k}$,
odd powers of $k^0$ or $\vec{k}$ vanish.
Then the expansion generates power series of
$(\Lambda_{\rm QCD}/(\alpha_S^2 m))^2$ and
$(\Lambda_{\rm QCD}/(\alpha_S m))^2$ in the power counting.
The order estimates of renormalons agree with Eq.~(\ref{suppressed}).

In the opposite case where the off-shellness is small,
$\alpha_S^2 m \ll \Lambda_{\rm QCD}$,
the power counting of Eq.~(\ref{potene}) leads to a different 
order estimate.
Since $k^0 \sim \frac{|\vec{k}|^2}{2m}$ or $\alpha_S^2m$, and
$\Bigl[(k^0)^2-|\vec{k}|^2 \Bigr] \sim \Lambda_{\rm QCD}^2$,
it follows that 
$ |\vec{k}|\sim  \Lambda_{\rm QCD} \gg
|k^0| \sim \frac{\Lambda_{\rm QCD}^2}{m}$ or 
$\alpha_S^2m $
(dominant contribution comes from an instantaneous gluon),
and the leading renormalon is estimated to be of order
$\Lambda_{\rm QCD}$.
Despite of this order estimate, the leading pole in
$\widetilde{E}_{{\rm pot},\beta_0}^{(1S)} [u]$
is located at  $u=1$ even for small off-shellness,
$0<\Delta \ll 1$,
and is not shifted to $u=1/2$.
To verify consistency, 
we now examine the uncertainty of the perturbative series of the
potential energy when the off-shellness is small.

First, by analyzing $I(u,\Delta)$ in the vicinity of $\Delta=0$,
we find that it can be decomposed into the parts 
which are analytic and non-analytic at $\Delta=0$:\footnote{
It is well known that asymptotic expansions of
loop integrals of Feynman diagrams 
consist of analytic and non-analytic parts in general.
See e.g.\ \cite{Larin01}.
}
\bea
I(u,\Delta) = I_{\rm A}(u,\Delta) + I_{\rm NA}(u,\Delta) .
\label{dec-I}
\eea
The analytic part coincides with $J(u)$ at $\Delta=0$:
\bea
I_{\rm A}(u,\Delta) = J(u) + {\cal O}(\Delta) .
\label{exp-A}
\eea
For $u>0$, the non-analytic part behaves as
\bea
I_{\rm NA}(u,\Delta) = \Delta^{1-2u}
\left[ \,
 I^{(0)}_{\rm NA}(u) + {\cal O}(\Delta)
\, \right] 
~~~~~~~~~~~
(\, u>0 \,) ,
\label{exp-NA}
\eea
where
\bea
&&
I^{(0)}_{\rm NA}(u)=
\frac{ 2 \,
         \Gamma(\frac{3}{2}+2u)
         \Gamma(1-u)
    }{\pi^{3/2}\, (u-\frac{1}{2})(u+\frac{1}{2})\,
         \Gamma(2+u) } .
\eea
$J(u)$ and $\Delta^{1-2u} I^{(0)}_{\rm NA}(u)$, respectively, 
have the leading poles at $u=1/2$ on the positive real axis.
These poles cancel with each other
(independently of the value of $\Delta$).
By contrast, in the vicinity of $u=0$, and if $\Delta \ll 1$, 
the contribution of
the non-analytic term is suppressed and $|I_{\rm A}| \gg |I_{\rm NA}|$ holds.

Let us write the contribution of the non-analytic term
$I^{(0)}_{\rm NA}(u)$ to 
$\widetilde{E}_{{\rm pot},\beta_0}^{(1S)} [u]$
(up to a normalization factor) as
\bea
x^{2u} \, \Delta^{1-2u} \, I^{(0)}_{\rm NA}(u)=\sum_n c_n(\Delta,x) \, u^n ,
~~~~~ ~~~~~
x \equiv \frac{\tilde{\mu}}{C_F\alpha_S(\mu)m} .
\eea
The asymptotic behavior of the expansion coefficient $c_n(\Delta,x)$
for large $n$ is determined by the pole and its residue at $u=1/2$
in the left-hand-side:
$
c_n(\Delta,x) \sim - \, {2^{n+2}} x / {\pi} .
$
As $\Delta$ becomes 
smaller, $c_n(\Delta,x)$ approaches this asymptotic behavior
more slowly.
Stating more precisely, the asymptotic form is a good approximation
if $n \simgt n_{x/\Delta} \equiv \log (x/\Delta)$.
When we estimate the uncertainty of the perturbative expansion of
the potential energy ${E}_{{\rm pot},\beta_0}^{(1S)}$, 
the size of the term in the vicinity of $n = n_\star$,
where it becomes minimal, matters.
If $n_\star \gg n_{x/\Delta}$ (i.e.\ $\alpha_S^2 m \gg \Lambda_{\rm QCD}$),
there are contributions from
both $I_{\rm A}$ and $I_{\rm NA}$ around $n = n_\star$;
the poles at $u=1/2$ cancel and the estimate of $I(u,\Delta)$
by the renormalon pole at $u=1$ is valid.
Oppositely, if $n_\star \ll n_{x/\Delta}$ 
(i.e.\ $\alpha_S^2 m \ll \Lambda_{\rm QCD}$),
the contribution from $I_{\rm NA}$ is suppressed around $n = n_\star$,
and the estimate by the $u=1/2$ pole of
$I_{\rm A}(u,\Delta) \approx J(u)$ is valid.
Thus, the behavior of the perturbative expansion at
$n \approx n_\star$ varies as if the position of the leading renormalon pole
is shifted, depending on the relative magnitude of
$\alpha_S^2 m$ and $\Lambda_{\rm QCD}$.

As we have seen, the leading renormalon contained in the potential
energy is suppressed when the off-shellness is large.
Then, an important question is whether 
the cancellation between the leading renormalons contained
in the potential energy $E_{\rm pot}$ and in 
the self-energies $E_{\rm SE}$ still takes place in this case.
Furthermore, if the cancellation occurs: 
Are the remaining sub-leading
renormalons in the total energy also suppressed as compared to those 
in $2 m_{\rm pole} + V_{\rm QCD}(r)$?
We consider these questions.
Using approximations consistent with those used for
${E}_{{\rm pot},\beta_0}^{(1S)}$,
we may estimate IR renormalons in the self-energy
contribution to the $1S$ energy level as
\bea
&&
{E}_{{\rm SE},\beta_0}^{(1S)} =
M - i \int \! \frac{d^4p}{(2\pi)^4} \, \,
|\phi_{C,1S}(\vec{p})|^2 \,
\left( \frac{1}{\frac{M}{2}+p^0-m-\frac{\vec{p}^{\,2}}{2m}+i\epsilon}
      +\frac{1}{\frac{M}{2}-p^0-m-\frac{\vec{p}^{\,2}}{2m}+i\epsilon}
\right)^2
\nonumber \\
&& ~~~~~ ~~~~~ ~~~~~ ~~~~~ 
\times
{\textstyle
\left[
\frac{M}{2}+p^0-m-\frac{\vec{p}^{\,2}}{2m} - \Sigma (p) - \delta m
\right] \,
\left[
\frac{M}{2}-p^0-m-\frac{\vec{p}^{\,2}}{2m} - \Sigma (-p) - \delta m
\right] 
}
\nonumber\\ 
&& ~~~~ ~~~~~
=
2 \, m + \frac{\Bigl( C_F \alpha_S(\mu) \Bigr)^2}{4} \, m
+ 2 \delta m +
E^{(1S)}_\Sigma
+ {\cal O}\Bigl((\Sigma+\delta m)^2\Bigr) ,
\label{ESEbeta0}
\eea
where
\bea
\Sigma (p) = i
\int \! \frac{d^4k}{(2\pi)^4}
\, \, \frac{1}{
\frac{M}{2}+p^0+k^0-m-\frac{|\vec{p}+\vec{k}|^{\,2}}{2m}+i\epsilon 
}
~~~~~~~~~~~~~~~~~~~~
\nonumber \\ 
\times \,
\frac{C_F \, 4 \pi\alpha_S(\mu) }
{| \vec{k} |^2}
\sum_{n=0}^\infty \left( \frac{\beta_0 \alpha_S(\mu)}{4\pi} \right)^n
\left\{ \log \Biggl( \frac{\tilde{\mu}^2}{-k^2} \Biggr) \right\}^n
\eea
represents the contribution of IR renormalons to the quark
self-energy.
The first two terms of Eq.~(\ref{ESEbeta0}) are the tree-graph 
contributions; 
$\delta m$ in the third term represents a scheme-dependent
mass counter-term (which is free from IR renormalons for a
short-distance mass);
the fourth term represents the
contribution from the self-energies of $Q$ and $\bar{Q}$
[linear term in $\Sigma$] induced by Coulomb gluon
in the large-$\beta_0$ approximation.
The Borel transform of $E^{(1S)}_\Sigma$ reads
\bea
&&
\widetilde{E}^{(1S)}_\Sigma [u] =
- \, \frac{\Bigl( C_F \alpha_S(\mu) \Bigr)^2 m}{2} \times
\left( \frac{\tilde{\mu}}{C_F \alpha_S(\mu) m} \right)^{2u} 
I_{\Sigma}(u,\Delta) ,
\label{Borelself} 
\eea
with
\bea
&&
I_{\Sigma}(u,\Delta)=
2^{8+2u} \, i \, \pi^2
\int \! \frac{d^3\vec{p}}{(2\pi)^3}
\int \! \frac{d^4k}{(2\pi)^4}
\, \, \, 
\frac{1}{
\Bigl( \vec{p}^{\,\,2}+1 \Bigr)^4 \,
\Bigl| \vec{k} \Bigr|^2 \,
\Bigl[ - (k^0 \Delta )^2 + |\vec{k}|^2 - i \epsilon \Bigr]^u
}
\nonumber \\
&&
~~~~~~~ ~~~~~~~~
\times 
\left(
\frac{1}{k^0+1+{\vec{p}^{\,\,2}}/{2}+{|\vec{p}+\vec{k}|^2}/{2}-i\epsilon}
+
\frac{1}{-k^0+1+{\vec{p}^{\,\,2}}/{2}+{|\vec{p}+\vec{k}|^2}/{2}-i\epsilon}
\right)
~~~
\label{Isig}
\nonumber\\
\\
&&
~~~~~~~~~~~~
=
-\frac{2^{3 + 2\,u} \,\Gamma({\frac{5}{2}} + u)\,\Gamma(2 + 2\,u)\, \Gamma(1 - u)
     }{3\,\pi^{3/2} \,\Delta \,\Gamma(u)\,\Gamma(3 + u)} 
\nonumber \\
&&
~~~~~~~~~~~~
~~~~
\times
\int_{0}^{1} d\xi \,\,\xi^{u-1}\, (1-\xi)^{\,3}\, (1+\xi)^{-2 u-3} 
{\textstyle
_{~2}F_{1} \! \left(\frac{5}{2}+u,1-u;3+u;1-\frac{\xi}{\Delta^2}\right)
}.
\label{oneparamintIsig}
\eea
It is easy to calculate the residue of 
$\widetilde{E}_{{\rm SE},\beta_0}^{(1S)} [u]$
at $u=N$ 
($N=1,2,3,\cdots$)
using the last line:
the residue is a $(2N-1)$-th-degree polynomial of 
$\Delta^{-1}$ and includes only odd powers of $\Delta^{-1}$
(similarly to the residue 
of $\widetilde{E}_{{\rm pot},\beta_0}^{(1S)} [u]$).
One may confirm that the order $\Delta^{1-2N}$ term of each residue
is equal in magnitude and opposite in sign to the corresponding term
of the residue of
$\widetilde{E}_{{\rm pot},\beta_0}^{(1S)} [u]$
at $u=N$.
Thus, the renormalons at order 
$\Lambda_{\rm QCD}\times ( \Lambda_{\rm QCD}/(\alpha_S^2 m))^{2N-1}$,
which are contained in the potential energy and the self-energy 
contribution, respectively,
get cancelled in the energy level 
$M_{\beta_0}^{(1S)} \equiv 
{E}_{{\rm SE},\beta_0}^{(1S)} + {E}_{{\rm pot},\beta_0}^{(1S)}$.
This means that all the uncertainties corresponding to $a=0$
in Eq.~(\ref{suppressed}) are cancelled.
The leading renormalon contained in $M_{\beta_0}^{(1S)}$ is of order
$\Lambda_{\rm QCD}
\times ( \Lambda_{\rm QCD}/(\alpha_S m))^2
\times ( \Lambda_{\rm QCD}/(\alpha_S^2 m))$.
Namely, renormalons contained in the energy level 
(when expressed in terms of a short-distance mass)
are suppressed by a factor $\Lambda_{\rm QCD}/(\alpha_S^2 m)$
as compared to those in 
$2 m_{\rm pole} + V_{\rm QCD}(r)$.
The orders of renormalons in $M_{\beta_0}^{(1S)}$ are consistent with
estimates based on power counting of the momentum-space integrals
(\ref{I}) and (\ref{Isig}).

It is necessary to investigate significance of the off-shell effects 
for the realistic heavy quarkonium states.
Let us consider the $1S$ states of the
(remnant of) toponium, bottomonium and charmonium, 
and set the quark masses $m$ as 175~GeV, 4.73~GeV($=M_{\Upsilon(1S)}/2$),
1.55~GeV($=M_{J/\psi}/2$), respectively.
Also, we choose the scales $\mu$ as
50~GeV, 2.49~GeV and 1.07~GeV, respectively.\footnote{
The scales for the bottomonium and charmonium $1S$ states are
taken from \cite{BraSuminoVai1};
the scale for the toponium $1S$ state is taken as (roughly) the
scale where the perturbative expansion becomes most convergent 
from the analyses of \cite{pro}.
}
Then the respective binding energies
$E_B = (C_F\alpha_S(\mu))^2 m /4$ are evaluated as
1.3~GeV, 0.16~GeV and 0.14~GeV.
We may consider them as typical sizes of off-shellness and compare
them with $\Lambda_{\rm QCD}=0.2$--0.3~GeV.
It seems certain that the off-shell effects are important in the
toponium energy level.
As for the bottomonium and charmonium states, more detailed
analyses may be needed to clarify the significance of the effects.
Here, as a simple analysis, we examine the series expansion of the
potential energy $E^{(1S)}_{{\rm pot},\beta_0}$
[Eq.~(\ref{PTexpansion})]
when we replace $I(u,\Delta)$ by $J(u)$ (on-shell approximation) and by
$J(u)+\Delta^{1-2u}I^{(0)}_{\rm NA}(u)$.
The terms up to $n=5$ are shown in Table~\ref{tab2}.
\begin{table}[t]
\begin{center}
\begin{tabular}{c|l|l|l|l|l|l|l|l}
\hline
\multicolumn{2}{c|}{} & 
\multicolumn{6}{c|}{
Order $(\beta_0\alpha_S)^n$ term of 
$E^{(1S)}_{{\rm pot},\beta_0}$ 
(LO in $\Delta$-expansion)
}&
\raise-10pt\hbox{$~~\Delta$} \\
\cline{3-8}
\multicolumn{2}{c|}{} &
$n=0$ & $n=1$ & $n=2$ & $n=3$ & $n=4$ & $n=5$ \\
\hline \hline
\raise-10pt\hbox{$t\bar{t}$} & A &
$-2.63$ & $-0.97$ & $-0.46$ & $-0.26$ & $-0.18$ & $-0.15$ & \raise-10pt\hbox{$0.087$} \\
& A+NA &
$-2.34$ & $-0.79$ & $-0.32$ & $-0.15$ & $-0.079$ & $-0.046$ &\\
\hline \hline
\raise-10pt\hbox{$b\bar{b}$} & A &
$-0.32$ & $-0.25$ & $-0.26$ & $-0.34$ & $-0.53$ & $-0.99$ & \raise-10pt\hbox{$0.18$} \\
& A+NA &
$-0.24$ & $-0.17$ & $-0.14$ & $-0.14$ & $-0.15$ & $-0.19$ &\\
\hline \hline
\raise-10pt\hbox{$c\bar{c}$} & A &
$-0.28$ & $-0.35$ & $-0.61$ & $-1.35$ & $-3.67$ & $-12.0$ & \raise-10pt\hbox{$0.30$} \\
& A+NA &
$-0.17$ & $-0.20$ & $-0.25$ & $-0.41$ & $-0.66$ & $-1.55$ & \\
\hline \hline
\end{tabular}
\end{center}
\caption{\footnotesize
The series expansion of the
potential energy $E^{(1S)}_{{\rm pot},\beta_0}$
[Eq.~(\ref{PTexpansion})]
when $I(u,\Delta)$ is replaced by $J(u)$ [A] and by
$J(u)+\Delta^{1-2u}I^{(0)}_{\rm NA}(u)$ [A+NA].
We choose
$\mu=50, ~2.49, ~1.07$~GeV, 
$m = 175, 4.73, 1.55$~{\rm GeV},
$n_l=5, ~4, ~3$ and
$\alpha_s^{(n_l)}(\mu)=0.130, ~0.274, ~0.448$, respectively,
for the toponium, bottomonium and charmonium $1S$ states.
}
\label{tab2}
\end{table}
Within this approximation, we find that
the off-shell effects cannot be neglected for all these states.
One should note, however, that,
since $\Delta$ are not particularly small for the bottomonium and
charmonium states (see the same table), errors due to
neglecting higher powers of $\Delta$ are
estimated naively to be 20--30\% for these states.

For a systematic analysis of the realistic quarkonium energy levels,
it is indispensable to calculate explicitly the first several terms
of the perturbative series of 
${E}_{{\rm pot},\beta_0}^{(1S)}$ and
${E}_{{\rm SE},\beta_0}^{(1S)}$.
The integral expression Eqs.~(\ref{oneparamint}) and (\ref{fxiu})
is not very suited for calculating the expansion coefficients of
$\widetilde{E}_{{\rm pot},\beta_0}^{(1S)} [u]$.
We will report the calculation of these coefficients and
their systematic analysis (together with the details of the
calculations of the present paper) in a future publication.

It has been known that the positions of renormalon poles 
are different between the observables constructed from the 
on-shell and off-shell Green functions \cite{Beneke99},
and that the orders of renormalons can be different in perturbative
QCD and in effective theories \cite{BenekeBraun94}.
In this paper we have shown, through an explicit calculation, 
how the orders of renormalons are
``shifted'' in the series expansion of the potential energy
when the off-shellness of $Q$ and $\bar{Q}$ is 
incorporated and varied.
We find the mechanism fairly interesting.
As far as we understand, all the previous perturbative analyses of heavy
quarkonium states do not incorporate the off-shell suppression effects.
This may originate from a failure to identify
all terms of the asymptotic expansion in $\Delta$ of
Eqs.~(\ref{exp-A}) and (\ref{exp-NA}), in particular
the non-analytic terms (cf.\ Ref.~\cite{ky-new}).
If we expand the integrand of Eq.~(\ref{I}) in $\Delta$ {\it before}
integration, some terms of the asymptotic expansion are lost.
We find that cancellation of renormalons in the total energy 
${E}_{{\rm pot},\beta_0} + {E}_{{\rm SE},\beta_0}$
is realized at a deeper level in perturbative QCD
than what has been known previously when
the off-shellness is large.
The remaining renormalon is of order 
$\Lambda_{\rm QCD}
\times ( \Lambda_{\rm QCD}/(\alpha_S m))^2
\times ( \Lambda_{\rm QCD}/(\alpha_S^2 m))$.

The dimension of this renormalon uncertainty 
is same as that of the non-perturbative correction
by the local gluon condensate of the Voloshin-Leutweyler type 
\cite{vol}.
We note that a priori
the two contributions are of qualitatively different nature.
The relation between renormalon uncertainties of perturbative
computations and non-perturbative contributions in operator product
expansion is studied in detail for the $R$-ratio \cite{Beneke99}.
As for the quarkonium energy level, 
Ref.~\cite{BraPineSotVai} discusses the relation between the order
$\Lambda_{\rm QCD}^3$ renormalon uncertainty in the static QCD potential
and (ultraviolet sensitivity of) a non-local gluon condensate within 
the pNRQCD framework; it concludes that
the former can be absorbed into the latter
(the perturbative renormalon is cancelled against
``the renormalon ambiguity of some non-perturbative effects'');
it does not calculate the expansion of the matrix element of the gluon 
condensate in $\alpha_S$, however.

\section*{Acknowledgements}

Y. K. was supported by the Japan Society for the Promotion
of Science.

\end{document}